\documentclass[twocolumn,showpacs,preprintnumbers,amsmath,amssymb]{revtex4}

\usepackage{dcolumn}
\usepackage{bm}
\usepackage{graphicx}

\begin{document}


\title{LISA, binary stars, and the mass of the graviton}

\author{Curt Cutler}
  \email{cutler@aei-potsdam.mpg.de}
  \affiliation{Max-Planck-Institut fuer Gravitationsphysik,
  Albert-Einstein-Institut, Am Muehlenberg 1,
  D-14476 Golm bei Potsdam, Germany}

\author{William A. Hiscock}
  \email{hiscock@physics.montana.edu}
\affiliation{Department of Physics \\ Montana State
University, Bozeman, MT 59717}%

\author{Shane L. Larson}
  \email{shane@srl.caltech.edu}
\affiliation{Space Radiation Laboratory \\ California Institute of
Technology, Pasadena, CA 91125}%

\date{\today}

\begin{abstract}
We extend and improve earlier estimates of the ability of
the proposed LISA (Laser Interferometer Space Antenna)
gravitational wave detector to place upper bounds
on the graviton mass, $m_g$, by comparing the arrival
times of gravitational and electromagnetic signals from
binary star systems. We show that the best possible limit
on $m_g$ obtainable this way is $\sim 50$ times better than
the current limit set by Solar System measurements. Among
currently known, well-understood binaries, 4U1820-30 is the best for
this purpose; LISA observations of 4U1820-30 should yield a
limit $\approx 3-4$ times better than the present
Solar System bound. AM CVn-type binaries offer the prospect of
improving the limit by a factor of $10$, {\it if} such systems can
be better understood by the time of the LISA mission.
We briefly discuss the likelihood that
radio and optical searches during the next decade will yield
binaries that more closely approach the best possible case.

\end{abstract}


\maketitle

Recent work by Will~\cite{Will} and Larson and Hiscock~\cite{LH} has
examined how gravitational wave observations by LIGO \cite{LIGO} and
LISA \cite{LISA} can be used to place upper bounds on the mass of the
graviton.  Will's proposed method utilizes the dispersion of the waves
generated in binary inspiral caused by a nonzero graviton mass, while
Larson and Hiscock have proposed direct correlation of gravitational
wave (GW) and electromagnetic (EM) observations of nearby white dwarf
binary star systems. Both approaches promise improved bounds
compared to the present bound based on Solar System dynamics, $m_g^{SS} < 4.4
\times 10^{-22}$ eV, which corresponds to a bound on the graviton Compton
wavelength of $\lambda_g^{SS} = h/m_g = 2.8 \times 10^{12}$ km \cite{Talmadge}.

The ultimate limit on the binary star method is determined by the
precision with which the phase of the gravitational wave signal can be
measured.  Larson and Hiscock~\cite{LH} estimated this uncertainty in phase by
considering the cadence of measurements made by LISA during a one-year
integration of the periodic signal from the binary star system.
However, this is not the dominant source of uncertainty in the
gravitational wave phase measurement.  Let $\phi$ be the orbital
phase of the binary at some fiducial time. For GW measurements with
signal-to-noise $S/N >> 1$, one can estimate the uncertainty with
which this phase can be extracted by
$\delta\phi_{GW} = \sqrt{(\Gamma^{-1})^{\phi\phi}}\bigl(1 + {\cal O}(S/N)^{-1}\bigr)$, where $\Gamma_{ij}$ is the Fisher information matrix.
For a circular binary, the signal is characterized by seven or eight physical
parameters: two angles describing the direction $\hat N$ to the binary (i.e., its
position on the sky), two more angles describing the normal $\hat L$
to the orbital plane, the overall amplitude $A$, the overall phase $\phi$,
the orbital
frequency $f_0$, and (non-negligible for for some binaries) the frequency
derivative $\dot f_0$.
If all parameters except $\phi$ are somehow already known
(e.g. if the binary is resolved optically, for example by the Space
Interferometry Mission (SIM) \cite{SIM}), then
$(\Gamma^{-1})^{\phi\phi} = (\Gamma_{\phi\phi})^{-1} =
\frac{1}{4}(S/N)^{-2}$ \cite{Cutler}, so we would estimate
$\delta \phi_{GW} \approx \frac{1}{2}\left( {S \over N} \right)^{-1}$.
More generally we can write
\begin{equation}
\delta \phi_{GW} = \frac{\alpha}{2}\left( {S \over N} \right)^{-1} \left(1 + {\cal O}(S/N)^{-1}\right)\;\;\; .
\label{phaselimit}
\end{equation}
\noindent where $\alpha (> 1) $ is a correction term that
accounts for the fact there will generally be {\it additional}
unknown parameters to be extracted from the GW data, which can
only increase $(\Gamma^{-1})^{\phi\phi}$. The case of interest,
for setting a limit on $m_g$, is where $\delta \phi_{GW} << 1$, or
${S \over N} >> 1$, so from now on we ignore the
${\cal O}(S/N)^{-1}$ correction in Eq.~(\ref{phaselimit}).

How large is $\alpha$ likely to be? To answer this question, we
must consider more carefully the definition of $\phi$, and
also consider what information will be available to supplement the
GW measurement.

By $\phi$, we will mean the orbital phase, at some fiducial time $t_0$,
measured from the point in the orbit where the orbital plane intersects the
plane perpendicular to the line-of-sight. More precisely,
\begin{equation}
\label{phidef1}
\phi \equiv {\rm tan}^{-1}\bigl[(\hat r(t_0) \cdot \hat y)/(\hat r(t_0)
\cdot \hat x)\bigr]
\end{equation}
\noindent
where $\hat r$ is the (unit) orbital separation vector (pointing from the
more massive to the less massive body),
$\hat x \equiv \hat N \times \hat L/||\hat N \times \hat L||$, and
$\hat y \equiv \hat L \times \hat x$. Fig. \ref{fig:orbit} illustrates these
quantities.
This is a useful definition when comparing
to most EM measurements, where
generally one does {\it not} know the overall orientation of the binary, but
{\it can} measure the phase of one body's motion towards or away from the
observer (e.g., by Doppler measurements, or, as discussed below
with 4U1820-30, because the flux is greatest at the instant the
the body is furthest from the observer). Rather than
report the phase at some fiducial time, astronomers typically
report some ``epoch''
(i.e., instant of time) $t_0$
when the orbital phase (measured optically) is $\phi_{EM} = 0$.
Of course, this optical (or X-ray, etc.) measurement will have some error;
we define $- \delta\phi_{EM}$ ($\approx - 2\pi f_0 \delta t_0$) to be the
true EM phase at time $t_0$.
Our proposal for constraining $m_g$ amounts to using
LISA to measure $\phi_{GW}(t_0)$. If photons and gravitons travel at
the same speed, then $\phi_{GW}(t_0)$ should be consistent with zero.
More precisely, one can set the following lower bound on the Compton
wavelength $\lambda_g$ of the graviton (see \cite{LH} for details):
\begin{equation}
   \lambda_g = \frac{1}{2 f_0} \sqrt{\frac{1}{2} \left(1+\frac{2 \pi f_o D}
    {\Delta} \right)} \; \; \; ,
   \label{comptonlimit1}
\end{equation}
where $D$ is the distance to the binary system
and $\Delta$ is defined by~\footnote{Unless otherwise noted, geometric units
where $G = c = 1$ are used throughout.}
\begin{equation}
\Delta = \sqrt{\delta \phi_{EM}^2 + \delta \phi_{GW}^2} \; \;
\; .
\label{Deltadefn}
\end{equation}
\noindent Note that in Eq.~(\ref{Deltadefn}) we have assumed that the
errors $\delta \phi_{EM}$  and $\delta \phi_{GW}$ are uncorrelated.

Eq.~(\ref{phidef1}) represents one convention for the
zero-point of the phase $\phi$; with a different convention,
Eqs.~(\ref{comptonlimit1})--(\ref{Deltadefn}) would remain valid, but with
different numerical values for $\delta \phi_{EM}$ and $\delta \phi_{GW}$.
Obviously, for any given binary, one wants to define $\phi$ in a way
that minimizes $\Delta$.  For instance, if the binary is seen nearly
face-on (i.e.,  $\hat L$ and $\hat N$ are nearly parallel), then
the $\phi$ defined in Eq.~(\ref{Deltadefn}) is difficult to measure, since
a small shift in $\hat L$ can produce a large shift in $\phi$,
for fixed $\hat r(t_0)$. In this case, if one could somehow resolve the
two components of the binary optically (e.g., with SIM), then one
would probably be better off choosing some arbitrary-but-easily-determined
direction (e.g., the direction to Polaris) to define
$\phi = 0$.

For the purpose of determining the best possible upper limit
one can set on $\lambda_g$, we will next assume that
$|\delta \phi_{EM}|$ is small compared to $|\delta \phi_{GW}|$.
We expect this will generally be the case when one can determine
the optical phase at all (because of the much higher $S/N$ one
typically has with optical measurements), but
this will have to be verified on a case-by-case basis.
Combining Eqs.~(\ref{phaselimit}), (\ref{comptonlimit1}),
and (\ref{Deltadefn}) with this assumption then allows us to
obtain a simple expression for the best possible limit achievable on
the Compton wavelength of the graviton from combined EM and GW
observations of binary systems:
\begin{equation}
\lambda_g \simeq {\pi\over {\sqrt{\alpha}}} \sqrt{D \over {2 \pi f_0}} \; \sqrt{S \over
N} \; \;
\; .
\label{comptonlimit}
\end{equation}

The rms $S/N$ (averaged over source locations and orientations)
is given by~\cite{Finn_Thorne}
\begin{equation}\label{sn}
S/N = h\, (2T)^{1/2}/\bigl[S_h^{SA}(f_{gw})\bigr]^{1/2} \, .
\end{equation}
\noindent where $h \equiv \sqrt{<h_+^2 + h_\times^2>}$ is the
rms value (averaged over time and source direction) of the
strain field at the detector, $T$ is the observation time,
$f_{gw} = 2 f_0$ is the gravity-wave frequency, and
$S_h^{SA}$ is the ``sky-averaged'' spectral density of the detector noise.
For a circular-orbit binary, $h = (\dot E_{gw})^{1/2}/(2\pi f_0 D)$, so the numerator in Eq.~(\ref{sn}) is given by
\begin{eqnarray}
    h \sqrt{2T}  &=&  7.52 \times 10^{-17}\, {\rm Hz}^{-1/2} \cdot
    \frac{M_1 M_2}{(M_1 +M_2)^{1/3}} \nonumber \\
    & \times & \left(\frac{100 \, {\rm pc}}{D}\right)
    \left(\frac{f_{0}}{10^{-3}\, {\rm Hz}} \right)^{2/3} \left(\frac{T}{1 \,
    {\rm yr}} \right)^{1/2}\ ,
    \label{SignalValue}
\end{eqnarray}
where the stellar masses $M_1$ and $M_2$ are measured in units
of the solar mass.
The sky averaged spectral density of detector noise is given
below the transfer frequency $f_{*} = c/(2 \pi L)$ by \cite{LHH1}
\begin{equation}
     S_{h}^{SA} = 2(S_{n}^{a} + S_{n}^{p})/R_{gw}\ ,
     \label{NoiseCurve}
\end{equation}
where $R_{gw} = 3/5$ is the low frequency gravitational wave
transfer function for LISA, and
\begin{equation}
     S_{n}^{a} = \frac{8 S_{a}} {L^{2} (2 \pi f_{gw})^{4} }\ ,
     \label{AccelPSD}
\end{equation}
and
\begin{equation}
     S_{n}^{p} = \frac {2 S_{x}} {L^{2}}\ ,
     \label{1wayPositPSD}
\end{equation}
are the spectral density of acceleration and position noise
(respectively) which are output through the detector.  $L$ is
the armlength of the interferometer.  The spectral densities
$S_a$ and $S_x$ are the raw spectral noise densities of the
noise acting on the detector.  The LISA specifications for these
values are $\sqrt{S_a} = 3 \times 10^{-15} {\rm m}/({\rm s}^2
\sqrt{\rm Hz})$ and $\sqrt{S_x} = 2.0 \times 10^{-11} {\rm
m}/\sqrt{\rm Hz}$ \cite{LISA},\footnote{Note that this value for
$S_x$ is specified for the {\it 1 way} position noise budget, as
is the expression in Eq.\ (\ref{1wayPositPSD}), and has many
constituent noises.  Often, pure shot noise is considered as the
limiting source of noise for the LISA floor; in that case, Eq.\
(\ref{1wayPositPSD}) should be replaced by Eq.\ (13) of
Reference \cite{LHH1}.}.

Since $S/N$ is inversely proportional to the distance to the binary, $D$,
$\lambda_g$ as given in Eq.(\ref{comptonlimit}) is actually independent of the
distance. (This independence of the limit on the source distance was
already noted by Will\cite{Will} in a similar context.)
It is then worthwhile to combine Eq.(\ref{comptonlimit}) and
Eq.(\ref{SignalValue}) to obtain
\begin{eqnarray}
\lambda_g & \simeq & 1.05 \times 10^4  \frac{\sqrt{M_1 M_2}}{(M_1 + M_2)^{1/6}}
     \left(\frac{10^{-3}\, {\rm Hz}}{f_o} \right)^{1/6} \nonumber \\
     &\times&  \alpha^{-1/2} \left(\frac{T}{1 \,
    {\rm yr}} \right)^{1/4}\left(S_h^{SA} \times ( 1 {\rm Hz})\right)^{-1/4}
    {\rm km} \;  .
\label{comptonlimit2}
\end{eqnarray}
Utilizing this expression for $\lambda_g$, and the low-frequency approximation to
the predicted LISA sensitivity curve as given by Eq.~({\ref{NoiseCurve}),
\footnote{The low frequency approximation is adequate for our purposes here.
If one uses the full sensitivity curve (available at
http://www.srl.caltech.edu/$\sim$shane/sensitivity/) instead, the results
obtained differ only insubstantially.},
we can now determine the best possible lower limit on
$\lambda_g$ that could be obtainable via this method, with an ``ideal'' binary
acting as the signal source. The system variables that appear in
Eq.~({\ref{comptonlimit2}) are the masses of the two stars, their orbital
frequency, and the sensitivity of the gravitational wave detector (``noise'',
to be evaluated at the frequency of the gravitational wave, $ f_{gw} = 2 f_o$).
The strongest bound on $\lambda_g$
will occur for a binary system whose orbital frequency $f_o$ is equal to
the frequency $f_c$ that minimizes the function $f^{2/3} S_{h}^{SA}(f)$.
Utilizing Eqs.~(\ref{NoiseCurve}-\ref{1wayPositPSD}), this minimum is found to
occur at a frequency
\begin{equation}
f_c \simeq 2.06 \times 10^{-3}\, {\rm Hz} \; \; ,
\label{critfrequency}
\end{equation}
and that the sky-averaged spectral density of detector noise at the corresponding
$f_{gw} = 2 f_c$ is
\begin{equation}
S_{h}^{SA}(2f_c) = 1.28 \times 10^{-40} \, {\rm Hz}^{-1} \; \; .
\label{optnoise}
\end{equation}

The strongest
limit on $\lambda_g$ is obtained by assuming that the stellar masses are equal
and as large as possible; we will take them both to be equal to the
Chandrasekhar mass, $ M_1 = M_2 \simeq 1.4 M_\odot$, which is the maximum mass
for a white dwarf, and appears to be the ``canonical'' mass for neutron stars,
based on observation. Evaluating $\lambda_g$ by substituting these values into
Eq.(\ref{comptonlimit2}), we find that for a 3-yr measurement, LISA could set
an upper limit of
\begin{equation}
\lambda_g^{\rm max} \simeq (\alpha^{-1/2}) 1.4 \times 10^{14}\, {\rm km} \; \; ,
\label{maxlambda}
\end{equation}
which is about $50/\sqrt{\alpha}$ times stronger than the present solar system limit on $\lambda_g$.
For more typical WD masses, $M_1 = M_2 = 0.5 M_\odot$, the improvement factor is still
$\approx 20/\sqrt{\alpha}$.

Note that in Eqs.~(\ref{critfrequency})--(\ref{maxlambda}) for the
optimum frequency and optimum limit, we have included only LISA's
instrumental noise. This is reasonable at gravity-wave frequencies
$f_{gw} \gtrsim 2$ mHz, but for $f_{gw} \alt 2$ mHz, confusion noise due
to unresolved galactic and extragalactic binaries is probably the
dominant LISA noise source, and the total noise increases 
(sum of instrumental and confusion noise) rises steeply
at lower frequencies. 
E.g., for frequencies $2f_0 \alt 1$ mHz (i.e, at GW frequencies a 
factor of two or more below where WD confusion noise 
begins to dominate), $f_0^{1/6} [S_{h}^{SA}(2f_0)]^{1/4} > 7 f_c^{1/6} [S_{h}^{SA}(2f_c)]^{1/4} $ 
(where here we include the WD confusion noise in $S_{h}^{SA}$).
Thus for binaries with $2f_0 \alt 1$ mHz, the best limit one could 
set on $m_g$ is only $\sim 7/\sqrt{\alpha}$ better than 
the Solar System
bound. For this reason, we will concentrate on binaries above the
frequency where confusion noise dominates:
$2f_0 \gtrsim 2$ mHz.

We now return to our discussion of $\alpha$.
For sources where an EM/GW comparison
can be made, it is clear that the
sky location $\hat N$ will be known to extremely high
accuracy from the EM signal. It is reasonable to expect the
frequency $f_0$ (at epoch $t_0$) and its derivative $\dot f_0$
can also be determined from the
EM data. That leaves four
parameters, including $\phi$, to be determined by LISA.
Following the methods described in~\cite{Cutler}, we
have calculated the Fisher matrix for this four-parameter
problem. We find that for the $\sim 20\%$ of cases
where $|\hat L \cdot \hat N| < 0.2$ (i.e., the cases where
the binary is seen nearly edge-on), the degradation of
$\delta \phi_{GW}$ is quite small: $\alpha < 1.2$.
And for the best $\sim 40\%$ of cases, the degradation factor
$\alpha$ is less than $\sim 3$. So a fair fraction
of sources will be favorably oriented for determining $\delta \phi_{GW}$.
Fortunately, the ``edge-on'' orientation that is favorable for
small $\delta \phi_{GW}$ is {\it also} favorable for small
$\delta \phi_{EM}$, since this orientation gives the largest
Doppler shift.

We next consider one source, 4U1820-30, which amounts to an ``existence
proof'' of the feasibility of this method of constraining $m_g$
by comparing optical and gravitational arrival times.
\subsection{4U1820-30}
Low-mass X-ray binary 4U1820-30 appears to consist of a low-mass
($\sim 0.07 M_\odot$) white dwarf in orbit around a NS.
The orbital period is $11.46 \pm 0.04$ min,
so $f_{GW} = 2.909$ mHz--i.e, a frequency where the galactic background can probably be subtracted out (so instrumental noise dominates), and close to the
optimum frequency for constraining $m_g$. The $11.46$ min periodicity
was first observed in X-rays, but was also recently detected in the
UV by the Hubble Space Telescope's Faint Object Spectrograph (FOS).
(The angular resolution of HST was required for the measurement,
since 4U1820-30 is in a very crowded field, near the core of
globular cluster NGC 6624.)
The UV modulation (and roughly its amplitude) had been predicted by
Arons \& King~\cite{Arons_King}, based on the following picture. The WD rotation
period is tidally locked to the orbital period, so that the same side always
faces the NS. This is the WD's ``hot side,'' as it is
heated by X-rays from the NS; the UV flux we measure varies as the hot side
is alternately facing towards and away from us. Clearly, the maximum
UV flux occurs when we see the hot side most nearly straight on, which
occurs at the point in the orbit when the NS is closest to us. This
observation provides crucial understanding of the relation of the phase
of the binary's light curve to that of the associated GW signal, which is
presently not understood for stronger and more well-known binary systems
such as AM CVn.
The measurements in Anderson et. al~\cite{Anderson} determined the
overall phase (equivalently, the epoch of UV maximum)
to within $\delta\phi_{EM} \sim 0.16$. (They state $\delta({\rm epoch}) =
0.0002$ days.)
Its distance is estimated at $7.6$ kpc, which
means LISA should detect it with
$S/N \approx 25$ (for a 3-yr observation, using the results from a single
synthesized Michelson). Based on Arons \& King~\cite{Arons_King},
the binary's inclination angle $i$ is estimated to be
$i \approx 43^\circ \pm 9^\circ$, but
this is somewhat model-dependent. We have calculated that $\alpha^{1/2}
\sim 4$ for binaries with $i \approx 45^\circ$, typically.
Using this value for $\sqrt{\alpha}$ in Eq.~(\ref{comptonlimit2}),
we estimate that LISA
observations of 4U1820-30 should improve the Solar System
bound on $m_g$ by a factor $\sim 3-4$. This improvement is comparable to
what should be obtainable by analysis of GW signals from
inspirals of stellar-mass compact objects observed by ground-based interferometers
such as LIGO and VIRGO \cite{Will}.
If we were to include constraints on the  allowed
range of $\hat L \cdot \hat N$
(from the optical measurements), when fitting to the GW data,
that would of course decrease $\alpha$ and so improve the limit.
\subsection{AM CVn-type binaries}
Several of the ``classic'' AM CVn-type binary systems offer high potential $S/N$
for gravitational wave observations, as well as having sufficiently short orbital
period to place their GW emission at frequencies high enough to avoid the confusion
noise from Galactic and extragalactic binaries. However, these Helium cataclysmic
variable systems, containing accretion disks, offer very complicated light curves that
make it difficult to understand how the binary's light curve phase is related to the
line of masses connecting the two stellar components. Unless the relative phase of the
EM and GW signals at the source is known, a binary cannot be used to place useful limits
on the graviton mass.

However, virtually all studies of these systems to date have utilized time-resolved
photometry; little or no time-resolved spectroscopic observations have yet been
dedicated to these systems. Time-resolved spectroscopic observations should be able to
provide Doppler information that will resolve the ambiguous relation between EM and GW
phases at the source. As an example, in the eponymous AM CVn system, the orbital
velocity of the primary star is about $40$ km/s; today, largely driven by the
extrasolar planet research efforts, Doppler surveys are reaching accuracies of
between $ 3-10$ m/s. If such accuracies can be obtained in spectrographic studies of
AM CVn-type binaries, the uncertainty in the EM phase, $\delta \phi_{EM}$, will be
significantly less than the uncertainty in the GW phase, $\delta \phi_{GW}$, for any
source for which $(S/N)_{GW} \alt 1000$. Since there is roughly a decade before LISA's
launch in which spectrographic techniques will continue to improve, and such observations
may be made of these binary systems, we feel there is a substantial chance their
nature may be sufficiently well understood so that their EM and GW signals may be used
to constrain $\lambda_g$.

In Table 1 we display some ``best limits'' that might be obtainable from the
higher frequency known AM CVn-type systems. The Table gives the period of
the binary system, along with the best lower limit on $\lambda_g$ that
might be attainable, and the ratio of that limit to the present solar system
bound on $\lambda_g$. In determining these ``best limits'', we have assumed
that optical astronomers will be able to adequately determine the physical
elements of these nearby binary systems, so that the primary limitation
on our method is the accuracy with which the phase of the gravitational wave
signal can be measured.
\begin{table}
    \centering
    \caption{Known AM CVn-type systems.}
    \begin{tabular}{lcccc}
     & Orbital Period & $\sqrt{\alpha}\lambda_g$ & $\sqrt{\alpha}\lambda_g/\lambda_g^{SS}$\\
     Name &  (s) & ($10^{12}$ km) & \\
      \tableline
    & & & & \\
    AM CVn       & $1028.73$ & $27.$        & $9.7$ \\
    EC15330-1403 & $1119$    & $23.$        & $8.1$ \\
    Cet3         & $620.26$  & $25.$        & $9.0$  \\
    RX J1914 +24 & $570$     & $36.$        & $13.$ \\
    RX J0806.3+1527 & $321$  & $25.$        & $8.9$ \\
  \end{tabular}
  \label{BinaryTable}
\end{table}

It is also worth noting that three of the five high-frequency systems listed here have
only recently been recognized as extremely short period binaries. Cet3 (also known
as KUV 01584-0939) was discovered in 1984, but its nature has been only just been
revealed by high speed photometry \cite{Warner}. Similarly for RX J1914+24
\cite{Ramsey,Wu} and RX J0806.3+1527 \cite{Israel}. This suggests that
additional such systems may well be discovered (or recognized as such) before
LISA's launch.
\subsection{Prospects for discovering a binary pulsar with $P_b \alt 1000$ s}
The discovery of a pulsar in a short-period ($P_b \alt 1000$ s) binary
with a NS or WD companion would likely provide an excellent system
for constraining $m_g$, for two reasons. First, a higher-mass system
tends to give a stronger limit on $m_g$. Second, relativistic corrections to
the binary orbit (perihelion precession and orbital inspiral)
and the pulse arrival time (the Einstein delay and Shapiro delay),
often allow most of the binary's parameters to be extracted.
In the best cases, all binary parameters are extracted, except for
one angle: the direction of $\hat L \times \hat N$.
So only two parameters, $\phi_{GW}$ and this direction angle, need to
be extracted from the GW data, which
should generally translate into a low $\alpha$.

The two shortest-period binary radio-pulsar systems currently known
(and also having companion mass $M_2 \ge 0.1 M_\odot$) are
are J0024-72W ($P_b = 2.6 {\rm hr}$, $M_2 = 0.15 M_\odot$, $d = 4.5$ kpc) and
B1744-24A ($P_b = 1.9 {\rm hr}$, $M_2 = 0.1 M_\odot$, $d = 7.1$ kpc)--
within a factor 16 and 12, respectively, of our ideal period $1/f_c = 0.16$ h.
(See Table 4 in Lorimer~\cite{Lorimer}.)
Recently discovered PSR J1141-6545 is also
notable in this context, because in addition to having a short-period
($P_b = 4.74$ hr), the mass of the companion WD is rather large:
$M_2 > 1.0 M_\odot$~\cite{Kaspi}.
Yungelson et al.~\cite{Yungelson} estimate that our Galactic disk
contains several tens of
NS's in binary systems with $f_0 > 1\,$ mHz ($\sim 40$ NS-WD's and
$\sim 10$ NS-NS's), so such short-period binary pulsars are likely to exist.
Until now there has been a strong selection effect against finding short-period
binary pulsars, since the orbital motion smears out the pulse
frequency, and the ``acceleration searches'' traditionally used to
demodulate the signal are ill-suited to observations
lasting longer than $\sim P_b/2\pi$. Significantly more sophisticated search
strategies are now being implemented (see Jouteux et al.~\cite{Jouteux} and references therein), so it is reasonable to expect a significant extension to
our database of binary pulsars, at the short-period end.

Even more promising are NS binaries in globular clusters.
Benacquista, Zwart, \& Rasio~\cite{Benacquista_et_al}
estimate that Galactic globular clusters contain
$\sim 2$ NS-NS and $\sim 10$ NS-WD binaries
(with a factor $\sim 10$ uncertainty in either direction)
with $f_0 > 1\,$mHz. Once LISA has detected these systems, LISA's
few-degree angular resolution should allow the host globular
cluster to be identified uniquely. And since LISA will provide
the orbital period and phase to high precision, there will be
only one orbital parameter to search over (the maximum velocity of the
NS along the line of sight), greatly facilitating radio identification
of any such sources that are first discovered by LISA.
\subsection{Conclusions}
We have shown that correlating EM and GW (LISA) observations of
LMXB 4U1820-30 should improve the current Solar System bound
on $m_g$ by a factor $\sim 3-4$.  We showed that for an ``ideal''
source, the improvement factor would be $\sim 50$. Since the
bound on $m_g$ that one obtains is independent of the distance to the source,
it seems almost inevitable that future EM surveys with
increased sensitivity will reveal new (generally more distant)
binaries that more closely approach the ideal improvement factor.
\begin{acknowledgments}
We thank M. Benacquista, R. Wade, and R. Hellings for helpful
conversations.  CC's work was partly supported by NASA grant
NAG5-4093.  The work of WAH was supported in part by National Science
Foundation Grant No.  PHY-0098787 and the NASA EPSCoR Program through
Cooperative Agreement No.  NCC5-579.  SLL acknowledges support for
this work under LISA contract number PO 1217163, and the NASA EPSCoR
Program through Cooperative Agreement NCC5-410.
\end{acknowledgments}

\newpage
\begin{figure}
\includegraphics{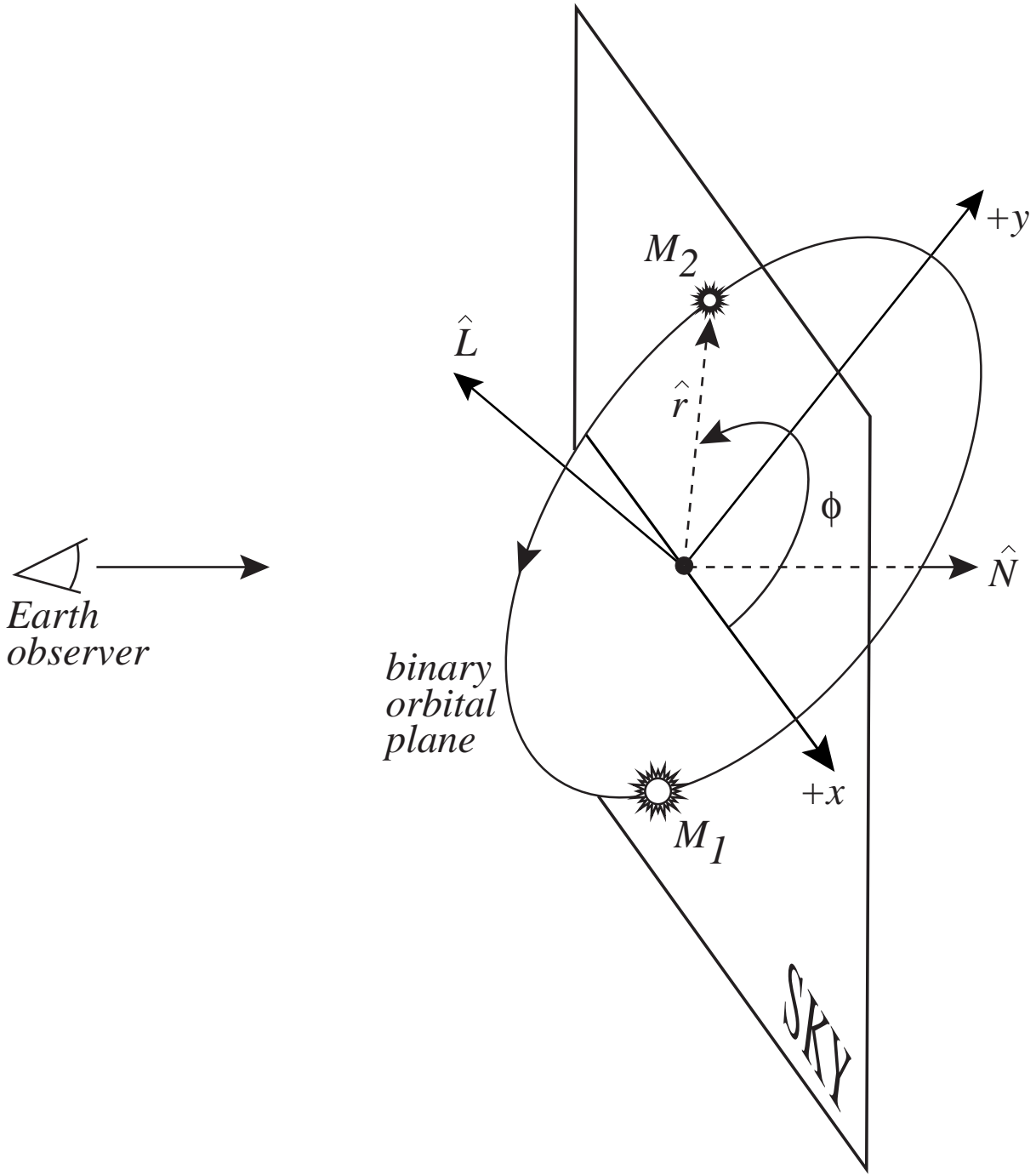}
      \caption{\label{fig:orbit}
      The coordinate system used to
      define the orbital phase is determined by the line of sight
      to the binary $\hat{N}$ and the binary orbital angular momentum
      $\hat{L}$, such that $\hat{x}$ points along the line of nodes,
      and $\hat{y}$ lies in the orbital plane.  The position vector
      $\hat{r}$ points along the binary axis, from the primary
      (largest mass , $M_{1}$) to the secondary (smallest mass,
      $M_{2}$) in the system.  The orbital phase $\phi$ is defined
      to be the angle in the orbital plane between the $+x$ axis
      and the position vector.}
\end{figure}
\end{document}